\documentclass[11pt]{article}

\usepackage{setspace,graphicx,epstopdf,amsmath,amsfonts,amssymb,amsthm,versionPO}
\usepackage{marginnote,datetime,enumitem,subfigure,rotating,fancyvrb}
\usepackage{hyperref,float}
\usepackage[longnamesfirst]{natbib}
\usepackage{algorithmic}
\usepackage{algorithm}
\usdate

\excludeversion{notes}		% Include notes?
\includeversion{links}          % Turn hyperlinks on?

\iflinks{}{\hypersetup{draft=true}}

\ifnotes{%
\usepackage[margin=1in,paperwidth=10in,right=2.5in]{geometry}%
\usepackage[textwidth=1.4in,shadow,colorinlistoftodos]{todonotes}%
}{%
\usepackage[margin=1in]{geometry}%
\usepackage[disable]{todonotes}%
}

\makeatletter\let\chapter\@undefined\makeatother % Undefine \chapter for todonotes

\usepackage{indentfirst} % Indent first sentence of a new section.
\usepackage{endnotes}    % Use endnotes instead of footnotes
\usepackage{jf}          % JF-specific formatting of sections, etc.
\usepackage[labelfont=bf,labelsep=period]{caption}   % Format figure captions
\begin{document}

\setlist{noitemsep}  % Reduce space between list items (itemize, enumerate, etc.)
\onehalfspacing      % Use 1.5 spacing
% Use endnotes instead of footnotes - redefine \footnote command
\renewcommand{\footnote}{\endnote}  % Endnotes instead of footnotes

%\author{RICHARD STANTON\thanks{\rm Stanton is with the Haas
      %School of Business, U.C. Berkeley. Acknowledgments \ldots}}

\title{\Large \bf Analysis of Secondary Effects in Roadside mmWave Backhaul Networks}
 
\author{\textit{Yuchen Liu} and \textit{Douglas M. Blough} \\
        School of Electrical and Computer Engineering \\ Georgia Institute of Technology}

\date{}              % No date for final submission

% Create title page with no page number

\maketitle
%\thispagestyle{empty}

%\bigskip

%\medskip

%\noindent JEL classification: XXX, YYY.

%\clearpage

% report start

\noindent{\large \bf Abstract}
\vspace{+0.2cm}

\noindent To achieve high survivability of mmWave wireless backhaul networks, deploying mmWave nodes on regularly-spaced lampposts in urban environment according to a triangular-wave topology is a promising approach, because the primary interference among the links on a path of mmWave nodes (referred to as self interference) can be eliminated, thereby maximizing end-to-end throughput. Besides, another advantage of this topology is the ability to reconfigure it to avoid obstacles that might occur along the roadway. Based on this network architecture, this work provides detailed analyses on the interference caused by secondary effects, which includes side-lobe effects and reflection effects. Through both analytical modeling and extensive evaluations, we show that the interference caused by secondary effects has only a very small impact on the network performance with the considered backhaul network architecture.

%In this part, we analyze the NLoS paths produced by reflection effects in our adopted roadside network scenarios, and we then conclude that 1) the interference caused by these NLoS paths have only a small impact on the network performance; 2) similar to LoS paths, the NLoS paths are also not available when the blockage occurs in most of cases.

% In this part, we analyze the secondary effects in our considered roadside network scenario, which includes potential interference due to side-lobe effects and ground, vehicle, or building reflections.

\section{Introduction}
With the advent of 5G and beyond, deploying mmWave small cell BSs along roadsides will be necessary to provide high data rate service to vehicles and their passengers.  As discussed in [1]-[3], these deployments will likely require the use of mmWave relays along the roadsides in between the small cell BSs.  A traditional ``straight-line'' topology, which mounts these relays on the tops of lampposts in a straight line, leads to poor system throughput due to severe self interference, and very limited ability to handle obstacles that block any of the links.  Therefore, the ``triangular-wave'' topology for relay-assisted backhaul is adopted, which is depicted in the blue links of Fig.~\ref{SL_Eff}.  In this topology, BSs and relays are deployed on equally-spaced lampposts on both sides of a road.  With equally-spaced lampposts, the angle $\theta$ between the center of a mmWave beam and the side of the road and the distance $d_{0}$  between the locations of consecutive nodes projected onto the same side of the road are the same everywhere along the topology.

%\vspace{-0.3cm} % adjust d of column btw fig. and front title 
%\begin{figure}[htbp]
%	%\vspace{-0.2cm} % adjust d of column btw fig. and front title
%	\setlength{\abovecaptionskip}{-0.2cm} % adjust d betw fig and title 
%	\centerline{\includegraphics[width=80mm]{road3.pdf}}
%	\caption{Topology model of relay-assisted mmWave backhaul.}
%	\label{triang_topol}
%\end{figure}
%\vspace{-0.1cm} % adjust d of column btw fig. and front title

One important feature of the triangular-wave topology is that with a large enough $\theta$ relative to the beamwidth of the mmWave directional antennas, self interference along the path is eliminated.  In our previous work, we proved that the following condition is necessary and sufficient to eliminate self-interference in the triangular-wave topology:
\vspace{-0.1cm} % adjust d of column btw fig. and front title
\begin{equation}
\theta  - \arctan (\frac{{\tan \theta }}{3}) > \frac{\phi }{2}\
\label{interf_free_eq}
\end{equation}
where $\phi$ is the beamwidth of the flat-top directional antennas used along the path. Because mmWave allows a high antenna density, narrow beamwidth directional antennas can be achieved through beamforming. As an example, with a beamwidth of $15^{\circ}$, Eq.~\ref{interf_free_eq} yields an angle $\theta$ of only around $12^{\circ}$. With such a small angle, the link length is increased by only a small amount compared to the straight-line topology but links become interference free, increasing the end-to-end throughput substantially.  In this paper, we investigate the secondary-interference effects in the triangular-wave topology. In what follows, we refer to this interference-free triangular-wave topology as IFTW.  Note that in the rest of this paper, we focus on the case where the data traffic is from left to right; however, the same reconfigured topology constructed from left to right will also work when the traffic direction is reversed.

With the standard assumption of additive white Gaussian noise channels, in following evaluations, the link capacity is assumed to follow Shannon's Theorem, and the rate of the directional unblocked link from node \textit{i} to node \textit{j} satisfies:

\begin{equation}
{R_{i,j}} \le \beta  \cdot B \cdot {\log _2}(1 + \min \{ \frac{{{P_r}(d)}}{{{N_T + I}}},{T_{\max }}\} )
\end{equation}

\noindent where \textit{B} is channel bandwidth, $P_{r}$ is the power of the intended transmitter's signal when it reaches the receiver, $N_{T}$ is the power of thermal noise, $T_{max}$ is the upper bound of operating signal-noise ratio because of the limiting factors like linearity in the radio frequency front-end, and the link utility ratio $\beta\in(0,1)$. Due to the primary interference, $\beta\le0.5$ and a maximum end-to-end throughput of more than 10 Gbps can theoretically be achieved in mmWave communications [8]. Because the considered topology is self-interference-free, here \textit{I} is the secondary interference from any transmitters, for example, caused by side-lobe emanation and reflection effects. And $P_{r}$ can be calculated as follow:

\begin{equation}
{P_r}(d) = {P_t} \cdot {G_t} \cdot {G_r} \cdot {(\frac{\lambda }{{4\pi d}})^\eta } \cdot {e^{ - \alpha d}}
\end{equation}

\noindent where $P_{t}$ is the transmit power, $G_{t}$ and $G_{r}$ are antenna gains at transmitter and receiver, respectively, $\lambda$ is the signal's wavelength, \textit{d} is the propagation distance, $\eta$ is the path loss exponent, and $\alpha$ is the attenuation factor due to  atmospheric absorption.
%Because of the severe path loss effect, we consider collective attenuation effects of dry air (including oxygen), water vapor, rain, and haze, so  
Here the small random attenuation caused by shadowing effect is ignored.

% In this work, a flat-top directional antenna model is considered, i.e. transceiver antennas have a high constant gain $G_{h}$ within the beam, and a very low gain $G_{l}$ that can be ignored outside the narrow beamwidth $\phi$.

\section{Analysis of Secondary Effects}

In this section, we analyze the secondary effects in the considered roadside network scenario, which includes the side-lobe effects and reflection effects.

\subsection{Side-lobe Effects}

\begin{figure}[hbtp]
	\vspace{-0.2cm} % adjust d of column btw fig. and front title
	\setlength{\abovecaptionskip}{+0.1cm} % adjust d betw fig and title 
	\centerline{\includegraphics[scale=0.8]{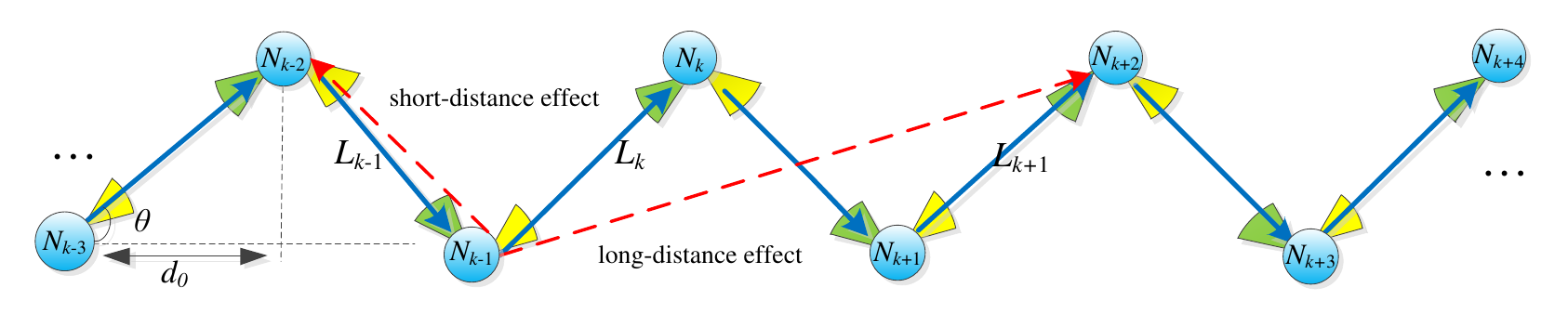}}
	\caption{Side-lobe effects in the IFTW topology.}
	\label{SL_Eff}
\end{figure}

Side-lobe effects indicates the interference caused by the antenna's side-lobe emanations. In the interference-free triangular-wave (IFTW) topology, we adopt the optimal scheduling, which contains two transmission slots of equal length with even numbered nodes transmitting in time slot 0 and odd numbered nodes transmitting in time slot 1. Therefore, for an arbitrary node $N_{k-1}$ (shown in Fig.~\ref{SL_Eff}), it may cause two kinds of side-lobe effects: short-distance effect and long-distance effect for $N_{k-2}$ and $N_{k+2}$, respectively. Note that we ignore the side-lobe effects on other further nodes (eg. $N_{k+4}$) due to the longer separation distances.  

\begin{figure}[hbtp]
	\vspace{-0.2cm} % adjust d of column btw fig. and front title
	\setlength{\abovecaptionskip}{+0.1cm} % adjust d betw fig and title 
	\centerline{\includegraphics[scale=0.75]{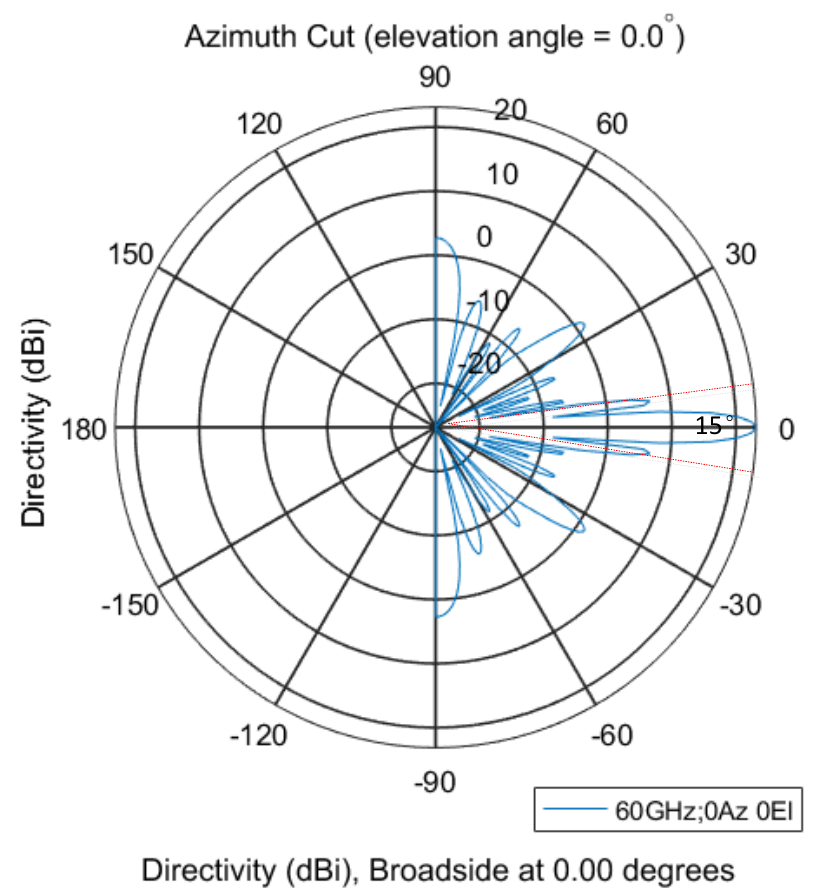}}
	\caption{Antenna directivity.}
	\label{Antenna}
\end{figure}

In this work, the narrow-beam directional antennas are adopted, and with the 61 element uniform hexagonal array antenna generated by Matlab software (shown in Fig.~\ref{Antenna}), the antenna gains of the mainlobe $G_h$ can be generated as 23.18 dBi and the sidelobe gain $G_l$ is lower than 2 dBi when the antenna beamwidth is around 15$^\circ$. Here given $G_l$ is 2 dBi, $d_0$ = 75m and $\theta$ = 11.7$^\circ$, we evaluate the SINR on receiver sides between the situations with/without side-lobe effects. As shown in Tab.~\ref{side-effect-comp}, we observe that the side-lobe effect has a very small impact on SINR (less than 0.6 dB), which can be ignored in the considered network scenarios. While this analysis applies to the topology prior to reconfiguration, any side-lobe effects that occur after reconfiguration will also have a small impact, for example, considering the worst case where the longer alternative link is reconfigured and both short- and long-distance effects occur at the receiver side, the impact on SINR is only around 0.5 dB with respect to the SINR of 25 dB under the normal situation.

 \begin{table}[htbp]
 	\centering
 	{\fontsize{9.5}{10.5}\selectfont
 		\caption{~SINR comparison between the situations with/without side-lobe effects.}
 		\label{side-effect-comp}
 		\begin{tabular}{|p{1.7cm}|p{1.9cm}|}	
 			%\begin{tabular}{|l|c|c|c|c|}
 			\hline
 			%\centering
 			\multicolumn{1}{|c|}{} & \multicolumn{1}{c|}{SINR on receiver side}	    \\
 			\hline
 			\multicolumn{1}{|c|}{Original (without side-lobe effect)} & \multicolumn{1}{c|}{41.1808 dB}	\\
 			\hline
 			\multicolumn{1}{|c|}{With short-distance effect} & \multicolumn{1}{c|}{40.5842 dB}	 \\
 			\hline
 			\multicolumn{1}{|c|}{With long-distance effect} & \multicolumn{1}{c|}{41.1731 dB}	 \\
 			\hline
 			%			\multicolumn{1}{|c|}{BTR(\%)} & \multicolumn{1}{c|}{98.46}	& \multicolumn{1}{c|}{93.21}   & \multicolumn{1}{c|}{98.46}           & \multicolumn{1}{c|}{N/A}    \\
 			%			\hline
 			%\end{tabular}
 		\end{tabular}
 	}
 \end{table}

\subsection{Reflection Effects}

As for reflection effects, we demonstrate that the interference due to reflections of the main beam and from side lobe emanations also has a small impact on network performance in our considered scenario. %can be ignored for two reasons: 1) the reflected interference only has a small impact on SINR when reflection occurs; 2) actually it has less possibility for receivers to receive reflection signals, because we adopt narrow-beam directional antenna (less than 15 degree) for each wireless node, and vehicles as main obstacles are usually moving fast, which might become the reflectors only if they parked in some particular places.
In what follows, we show the analysis of reflection effects in detail.

\subsubsection{Vehicle-reflection Effects}

\begin{figure}[hbtp]
	\vspace{-0.2cm} % adjust d of column btw fig. and front title
	\setlength{\abovecaptionskip}{+0.1cm} % adjust d betw fig and title 
	\centerline{\includegraphics[scale=0.8]{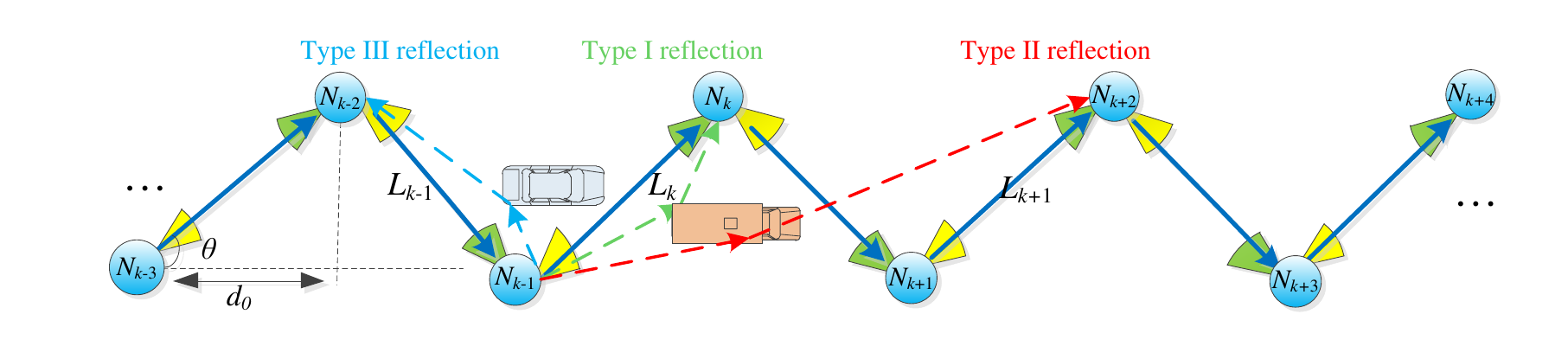}}
	\caption{Three types of reflection in the IFTW topology.}
	\label{re-eff}
\end{figure}

First, considering the primary interference constraint and two-slot optimal scheduling, there are three types of vehicle reflection that might occur in the topology, which is shown in Fig.~\ref{re-eff}.

The \textit{Type} I reflections (see the green dashed line in Fig.~\ref{re-eff}) cause a multi-path environment, which can be helpful in increasing the data rate of the link if the receiver $N_k$ accounts for this extra diversity. Thus, this reflection can benefit the link performance instead of producing interference.  In what follows, we focus on destructive interference caused by \textit{Type} II and \textit{Type} III reflections.

As for the \textit{Type} II reflection (see the red dashed line in Fig.~\ref{re-eff}), the transmission signal from node $N_{k-1}$ could be possibly reflected from the roof of vehicle and become interference at $N_{k+2}$. Assuming the tops of vehicles are flat that will not change the signal's direction, and based on the interference-free feature of topology, this interference signal would be emanated  and received by the antenna's side lobes of $N_{k-1}$ and $N_{k+2}$, respectively, and the longer separation distance would also make the effect of this interference become fairly smaller. Considering the body of vehicles as reflectors, where the reflection loss of metal materials are 0 dB [4], we evaluate the SINR between the situations with/without the \textit{Type} II reflection. As shown in Tab.~\ref{TypeII-comp}, we observe that it only causes less than 0.1 dB loss of SINR on the receiver side if \textit{Type} II reflection occurs.

\begin{table}[htbp]
	\centering
	{\fontsize{9.5}{10.5}\selectfont
		\caption{~SINR comparison between the situations with/without \textit{Type} II reflection.}
		\label{TypeII-comp}
		\begin{tabular}{|p{1.7cm}|p{1.9cm}|}	
			%\begin{tabular}{|l|c|c|c|c|}
			\hline
			%\centering
			\multicolumn{1}{|c|}{} & \multicolumn{1}{c|}{SINR on receiver side}	    \\
			\hline
			\multicolumn{1}{|c|}{Original (without reflection)} & \multicolumn{1}{c|}{41.1808 dB}	\\
			\hline
			\multicolumn{1}{|c|}{With reflection} & \multicolumn{1}{c|}{41.1768 dB}	 \\
			\hline
			%			\multicolumn{1}{|c|}{BTR(\%)} & \multicolumn{1}{c|}{98.46}	& \multicolumn{1}{c|}{93.21}   & \multicolumn{1}{c|}{98.46}           & \multicolumn{1}{c|}{N/A}    \\
			%			\hline
			%\end{tabular}
		\end{tabular}
	}
\end{table}

For the \textit{Type} III reflection shown with the blue dashed line in Fig.~\ref{re-eff}, the reflected signal from the side lobe of antenna of node $N_{k-1}$ will become the interference signal of $N_{k-2}$, and through the evaluation (see Tab.~\ref{TypeIII-comp}), it is observed that the reduction of SINR caused by \textit{Type} III reflection is less than 0.4 dB.

\begin{table}[htbp]
	\centering
	{\fontsize{9.5}{10.5}\selectfont
		\caption{~SINR comparison between the situations with/without \textit{Type} III reflection.}
		\label{TypeIII-comp}
		\begin{tabular}{|p{1.7cm}|p{1.9cm}|}	
			%\begin{tabular}{|l|c|c|c|c|}
			\hline
			%\centering
			\multicolumn{1}{|c|}{} & \multicolumn{1}{c|}{SINR on receiver side}	    \\
			\hline
			\multicolumn{1}{|c|}{Normal (without reflection)} & \multicolumn{1}{c|}{41.1808 dB}	\\
			\hline
			\multicolumn{1}{|c|}{With reflection} & \multicolumn{1}{c|}{40.7836 dB}	 \\
			\hline
			%			\multicolumn{1}{|c|}{BTR(\%)} & \multicolumn{1}{c|}{98.46}	& \multicolumn{1}{c|}{93.21}   & \multicolumn{1}{c|}{98.46}           & \multicolumn{1}{c|}{N/A}    \\
			%			\hline
			%\end{tabular}
		\end{tabular}
	}
\end{table}

As a result, the interference due to reflection effects caused by vehicles has a negligible impact on SINR ($\sim$1 dB in total) in the considered network scenario. Any reflection effects that occur after reconfiguration will also have a small impact, since even in the worst case, where a longer alternative link is reconfigured with the weaker signal at receiver side, there is only less than 1 dB reduction of SINR caused by reflection effects with respect to the SINR of 25 dB in the normal situation.

\subsubsection{Building-reflection Effects}

\begin{figure}[hbtp]
	\vspace{-0.1cm} % adjust d of column btw fig. and front title
	\setlength{\abovecaptionskip}{+0.1cm} % adjust d betw fig and title 
	\centerline{\includegraphics[scale=0.8]{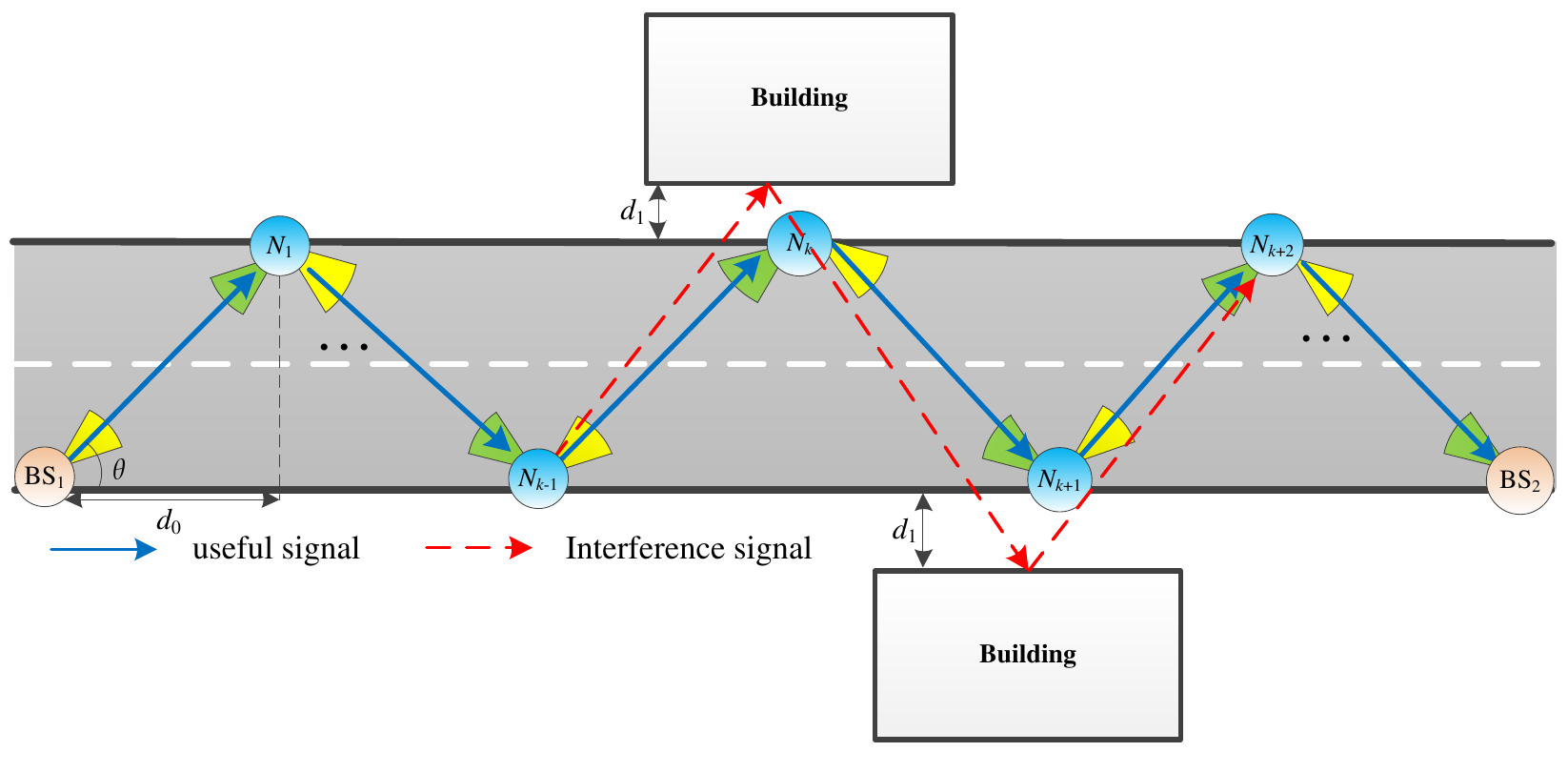}}
	\caption{Building-reflection effect in the IFTW topology.}
	\label{Build-reff}
\end{figure}

In the roadside-deployment scenarios, we also consider a possible situation where some buildings behind the lampposts may reflect the signal that could produce the secondary-interference. As shown in Fig.~\ref{Build-reff}, the signal from $N_{k-1}$ will interfere the signal receiving of $N_{k+2}$ through double reflections. Note that $N_{k+1}$ will not be interfered by the first reflection due to the two-slot optimal scheduling [3].    

\begin{figure}[hbtp]
	\vspace{-0.2cm} % adjust d of column btw fig. and front title
	\setlength{\abovecaptionskip}{+0.1cm} % adjust d betw fig and title 
	\centerline{\includegraphics[scale=0.7]{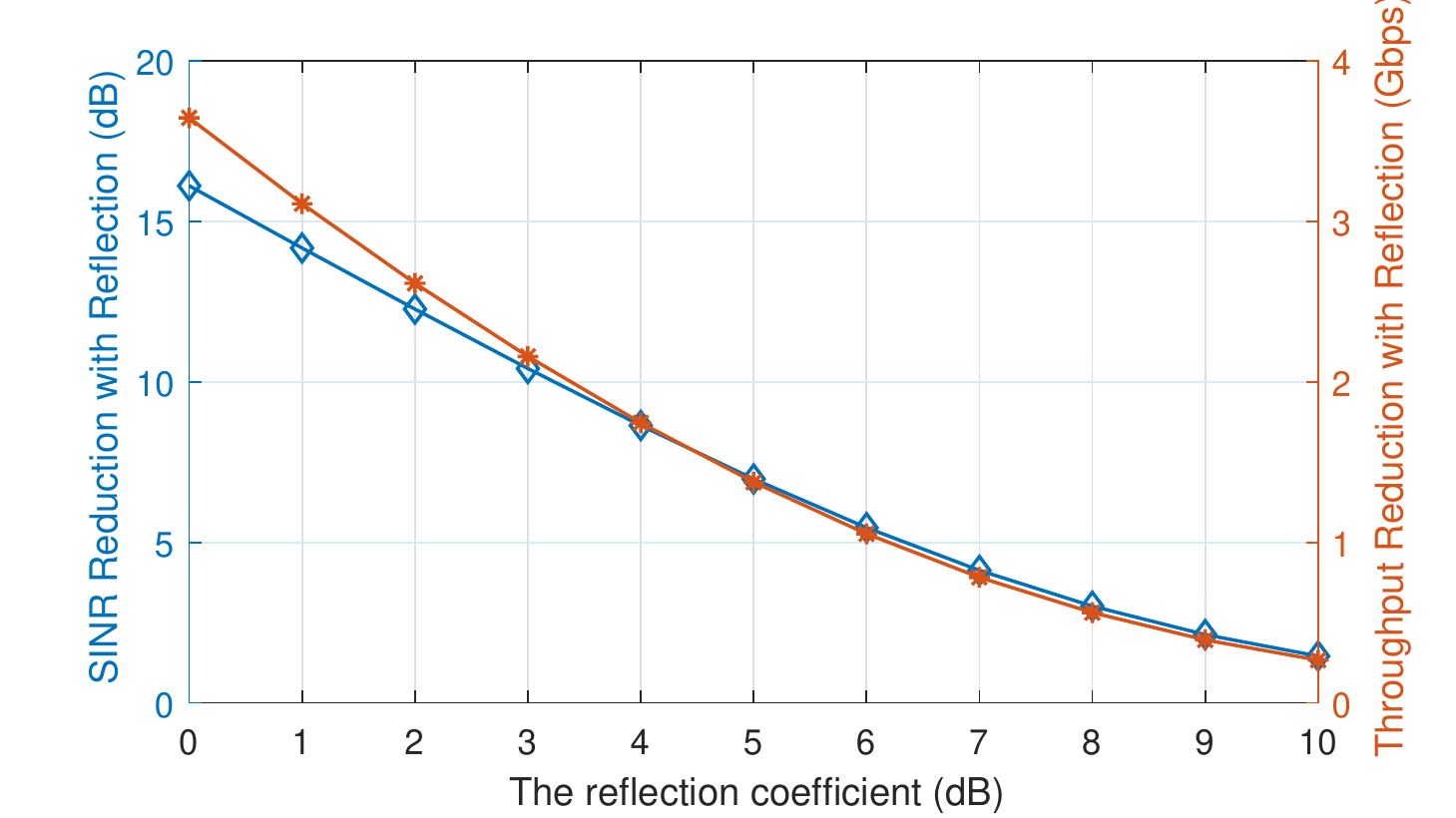}}
	\caption{Performance loss caused by building-reflection effects vs. reflection coefficient.}
	\label{build_gamma}
\end{figure}

Here we consider the worst case, where the building is very close to the lamppost, and $N_{k+2}$ receives the interference signal from $N_{k-1}$ (actually it would hardly be received due to the strict geometry constraint and the narrow beam of adopted antenna). First, given $d_1$ is 4m, Fig.~\ref{build_gamma} shows how performance loss varies in terms of reflection coefficient $\Gamma$. According to the measurement from [4]-[7], the reflection coefficients of some building materials are shown as: concrete is 7.5 dB, brick is 14.8 dB and glass is 8 dB, therefore, assuming $\Gamma$ is around 8 dB, we observe the reduction of SINR is only around 3 dB.

\begin{figure}[hbtp]
	\vspace{-0.2cm} % adjust d of column btw fig. and front title
	\setlength{\abovecaptionskip}{+0.1cm} % adjust d betw fig and title 
	\centerline{\includegraphics[scale=0.7]{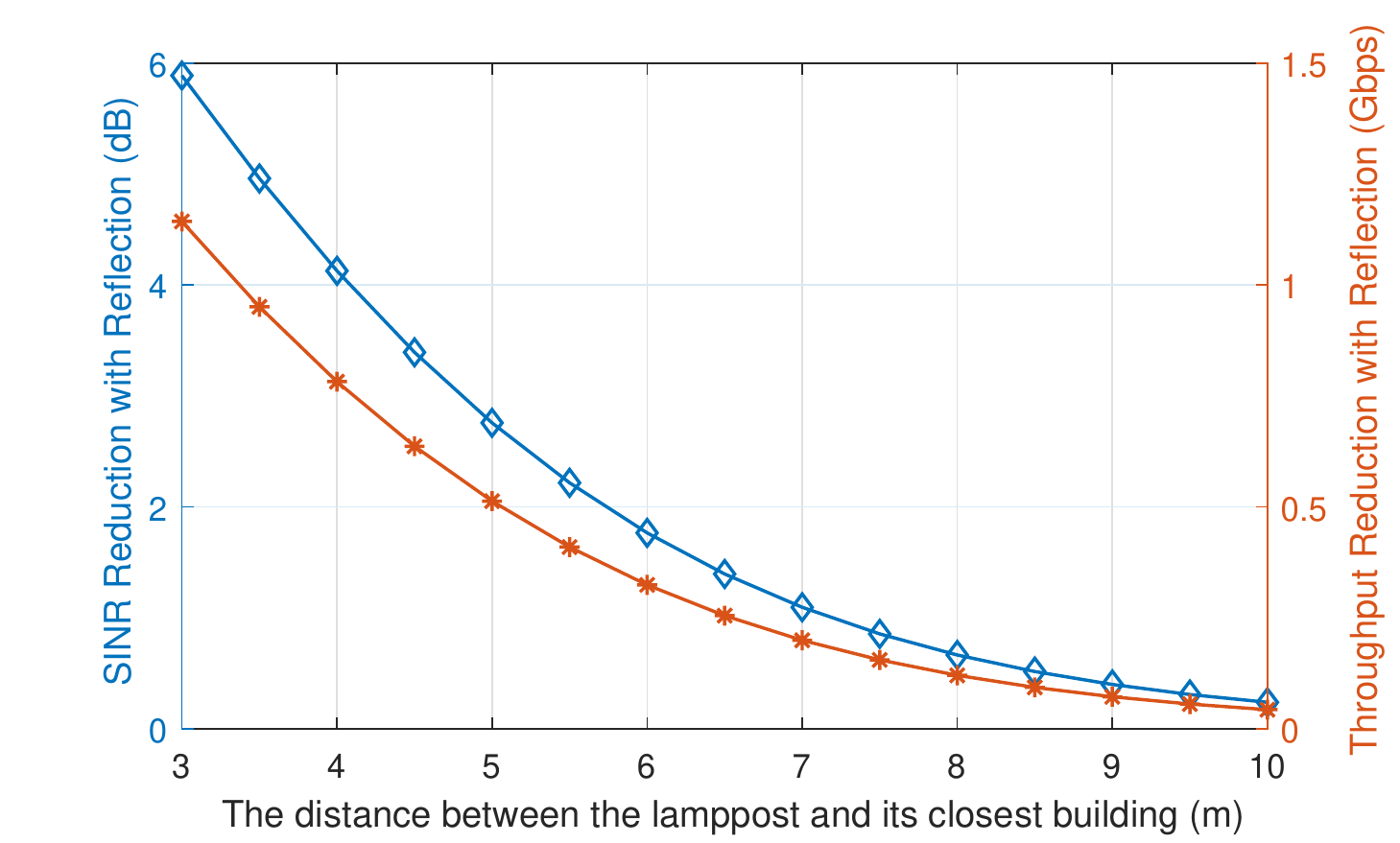}}
	\caption{Performance loss caused by building-reflection effects vs. $d_1$.}
	\label{build_distance}
\end{figure}

Second, given $\Gamma$ is 7 dB, Fig.~\ref{build_distance} shows how performance loss varies with the building-lamppost separation distance $d_1$. By measuring $d_1$ from the Google map in Manhattan, the distance between lamppost and its behind building is around 4m, thus we find that the reduction of SINR caused by this building-reflection effect is around 3 dB. In addition, we observe that when $d_1$ is larger than 7m, the reduction of SINR will be typically less than 1 dB.

\begin{figure}[hbtp]
	\vspace{-0.2cm} % adjust d of column btw fig. and front title
	\setlength{\abovecaptionskip}{+0.1cm} % adjust d betw fig and title 
	\centerline{\includegraphics[scale=0.7]{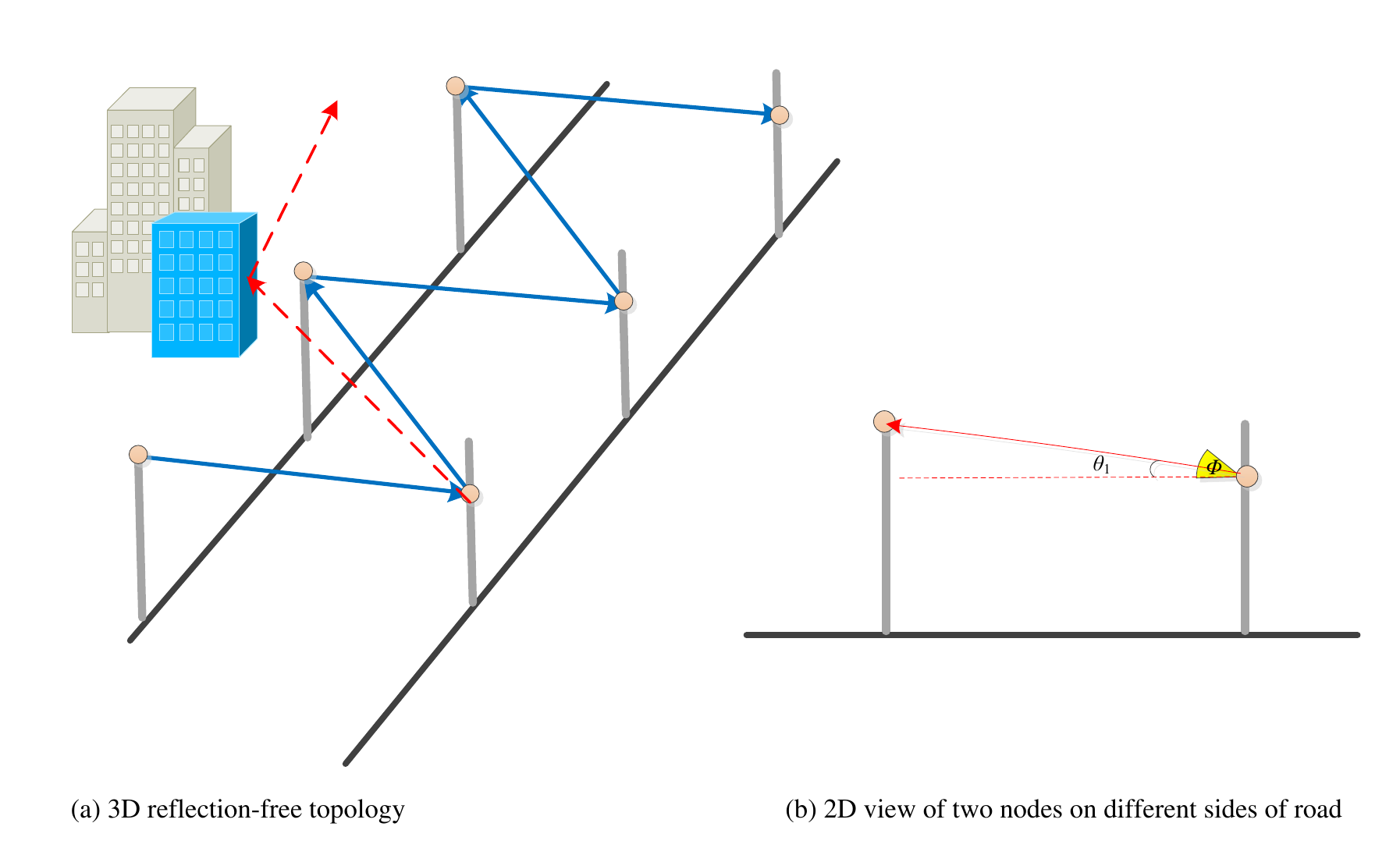}}
	\caption{The approach to eliminate building-reflection effects.}
	\label{build_3D}
\end{figure}

While considering some rare cases, if two buildings are made up of metals and located at exactly right spots, the $\gamma$ is around 0 dB and the average reflection loss would become fairly small, therefore, the performance loss caused by the building-reflection effects could be an issue (see Fig.~\ref{build_gamma}). However, we can eliminate this reflection effect by slightly adjusting our topology (shown in Fig.~\ref{build_3D}), i.e., slightly decrease the height of nodes on one side of lampposts, causing the signal at most reflect for once and finally go to the sky when $\theta_1 \le \phi/2$ (see Fig.~\ref{build_3D} (b)). For example, if the antenna's beamwidth $\phi$ is 8$^\circ$ and the width of road is 10m, by decreasing the height of relays on one side to 2.8m (original height is 3.5m), this building-reflection effect can be completely eliminated.

\subsubsection{Reflection Probability}
As we know, the building-reflection effects would not be an issue with the presented approach in Fig.~\ref{build_3D}, but the vehicle-reflection effects could also occur, therefore, here we derive the occurrence probability of interference caused by vehicle-reflection effects. In this work, we focus on the analysis of $Type$ II vehicle reflection, and the same analytical approach can be used on other reflection types, which will produce the similar results. %Since we can adjust topology to eliminate the building-reflection effects, here only $Type$ II reflection effect (see Sec. B) is considered.

\begin{figure}[hbtp]
	\vspace{-0.2cm} % adjust d of column btw fig. and front title
	\setlength{\abovecaptionskip}{+0.1cm} % adjust d betw fig and title 
	\centerline{\includegraphics[scale=0.45]{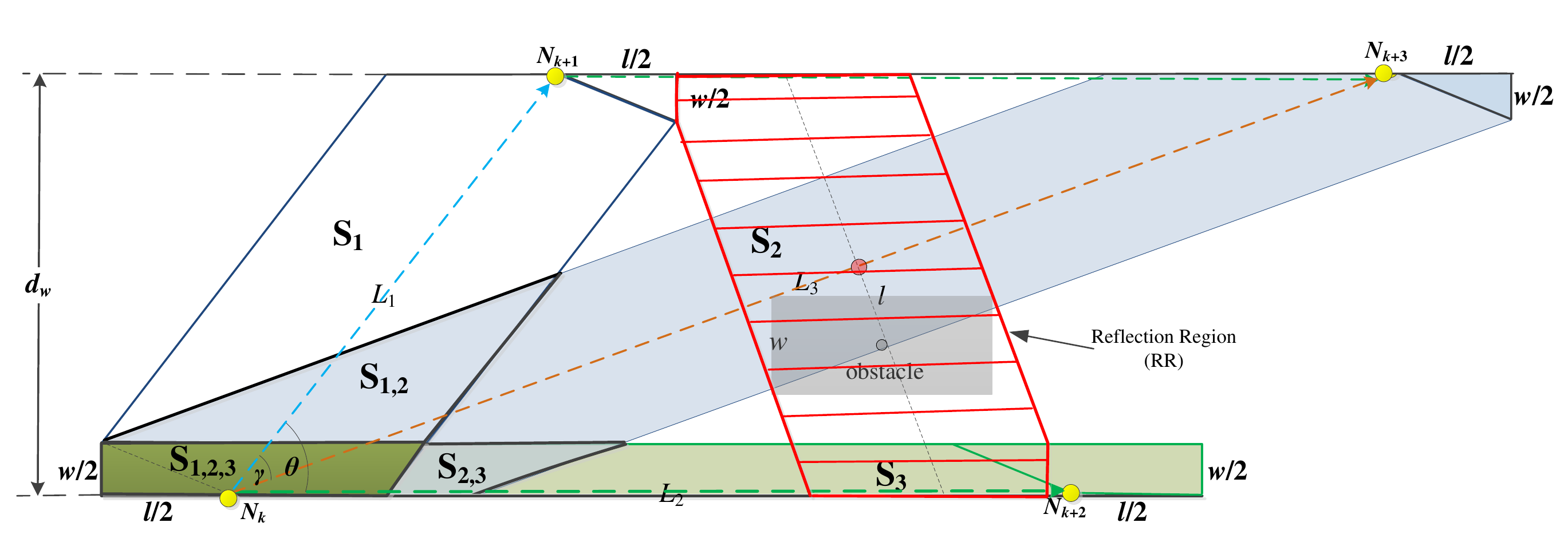}}
	\caption{Reflection area in one link region.}
	\label{Refl_area}
\end{figure}

Fig.~\ref{Refl_area} shows the top view of one link region in the IFTW topology. Assuming the widths and lengths of random-located vehicles are normally distributed as $\mathcal{N}$($\mu_w$=2.3, $\sigma_w$=1.2) and $\mathcal{N}$($\mu_l$ =5.5, $\sigma_l$ =3.5), respectively, and their centers follow homogeneously Poisson point process (PPP) with density $\lambda$ . By some geometric analysis, we first derive the area of reflection region (RR) as:

\begin{equation}
 \begin{array}{l}
 {\Lambda _{RR}} = \int_W {\int_L {\lambda  \cdot {S_{RR}}(w,l,\theta )} }  \cdot {f_L}(l) \cdot {f_W}(w)dwdl \\ 
 ~~~~~~{\rm{      }} = \lambda  \cdot d \cdot \tan \theta  \cdot {\mu _l} + \lambda  \cdot \tan (\theta  - \gamma ) \cdot \tan \theta  \cdot {\mu _w} - \frac{1}{4} \cdot \lambda  \cdot \tan (\theta  - \gamma ) \cdot ({\mu _w}^2 + {\sigma _w}^2) \\ 
 \end{array}
\end{equation}

Considering a $N$-node IFTW topology, there are N-3 non-overlapping RRs, thus we can get the reflection probability as: ${P_{R}} = 1- \exp (-E[K]) = 1 - \exp [ - (N - 3) \cdot {\Lambda _{RR}}]$ according to the PPP proposition, where $K$ is the total number of vehicles falls in RRs. If we incorporate the height of vehicles in this mathematical model and assume that the height of vehicles is also normally distributed as $\mathcal{N}$($\mu_h$ = 3.0, $\sigma_h$ = 1.5), the expected number of vehicles falls in RR is also a Poisson random variable with E[$K'$] = $\eta$E[$K$], where 

\begin{equation}
\eta  = \int\limits_{{h_r} - {h_{refl}}}^{{h_r}} {{f_H}(h)dh}  = \int\limits_{{h_r} - \frac{1}{2} \cdot d \cdot \tan (\frac{\phi }{2}) \cdot \sqrt {{{\tan }^2}\theta  + 9} }^{{h_r}} {\frac{1}{{\sqrt {2\pi } {\sigma _h}}} \cdot \exp ( - \frac{{{{(h - {\mu _h})}^2}}}{{2{\sigma _h}^2}})dh}
\end{equation}

\noindent In this way, we obtain the reflection probability in the entire network topology as:

\begin{equation}
{P_{Refl}} = 1 - \exp [ - (N - 3) \cdot \eta  \cdot {\Lambda _{RR}}]
\end{equation}

\begin{figure}[hbtp]
	\vspace{-0.2cm} % adjust d of column btw fig. and front title
	\setlength{\abovecaptionskip}{+0.1cm} % adjust d betw fig and title 
	\centerline{\includegraphics[scale=0.62]{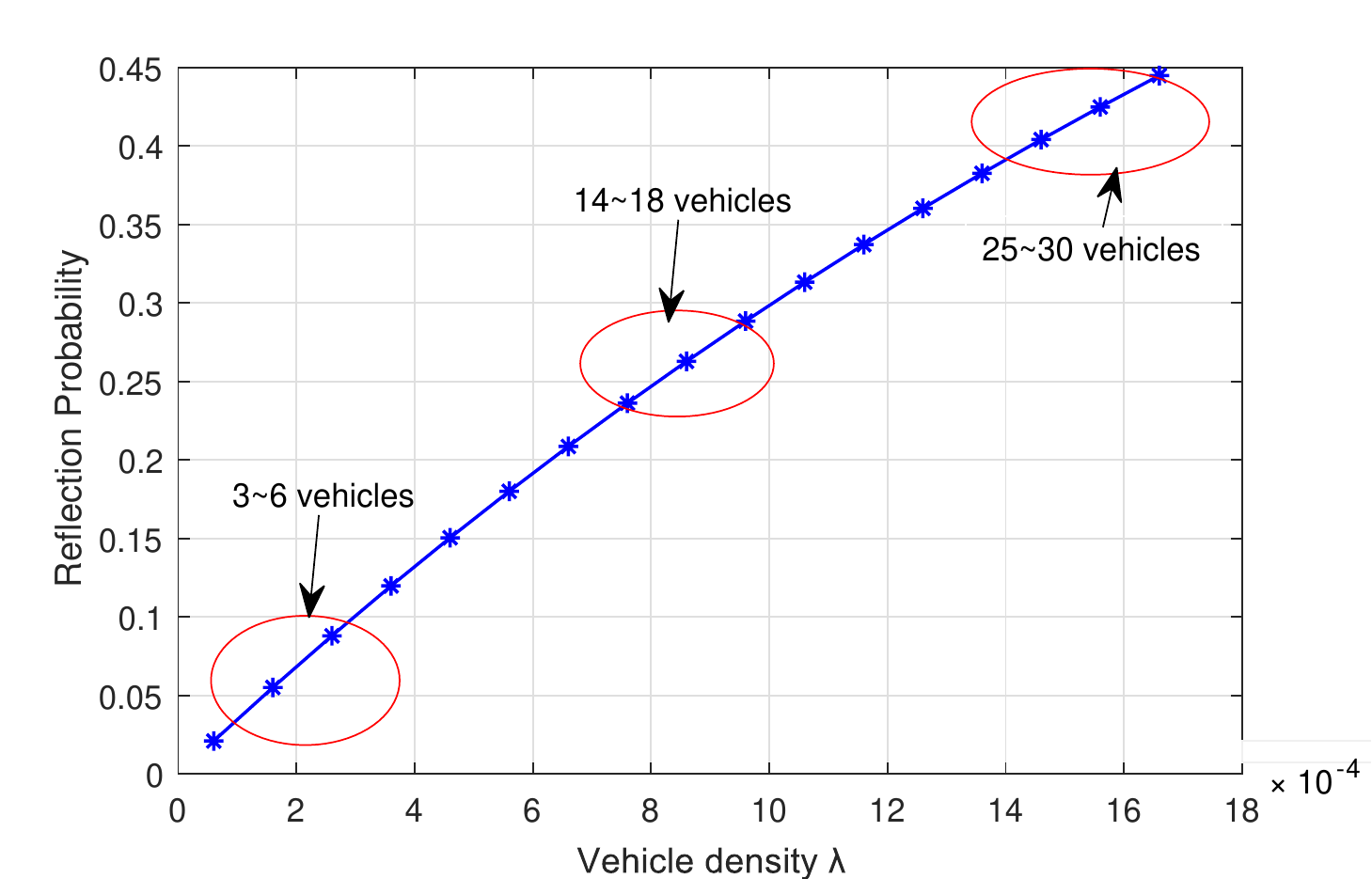}}
	\caption{Reflection probability vs. vehicle density $\lambda$.}
	\label{Refl_Prob}
\end{figure}

Considering a 10-node topology on the road (about 1000m), we evaluate the reflection probability vs. the vehicle density. As Fig.~\ref{Refl_Prob} shows, when the traffic on the road is not intense, i.e. with few vehicles, the reflection probability is less than 10\%. Under the intense-traffic situation, where there are around 15 vehicles on the road, the reflection probability is around 25\%. %In the worse cases, there are 30 cars on the road, the reflection probability is still less than 45\%.

\begin{figure}[hbtp]
	\vspace{-0.2cm} % adjust d of column btw fig. and front title
	\setlength{\abovecaptionskip}{+0.1cm} % adjust d betw fig and title 
	\centerline{\includegraphics[scale=0.7]{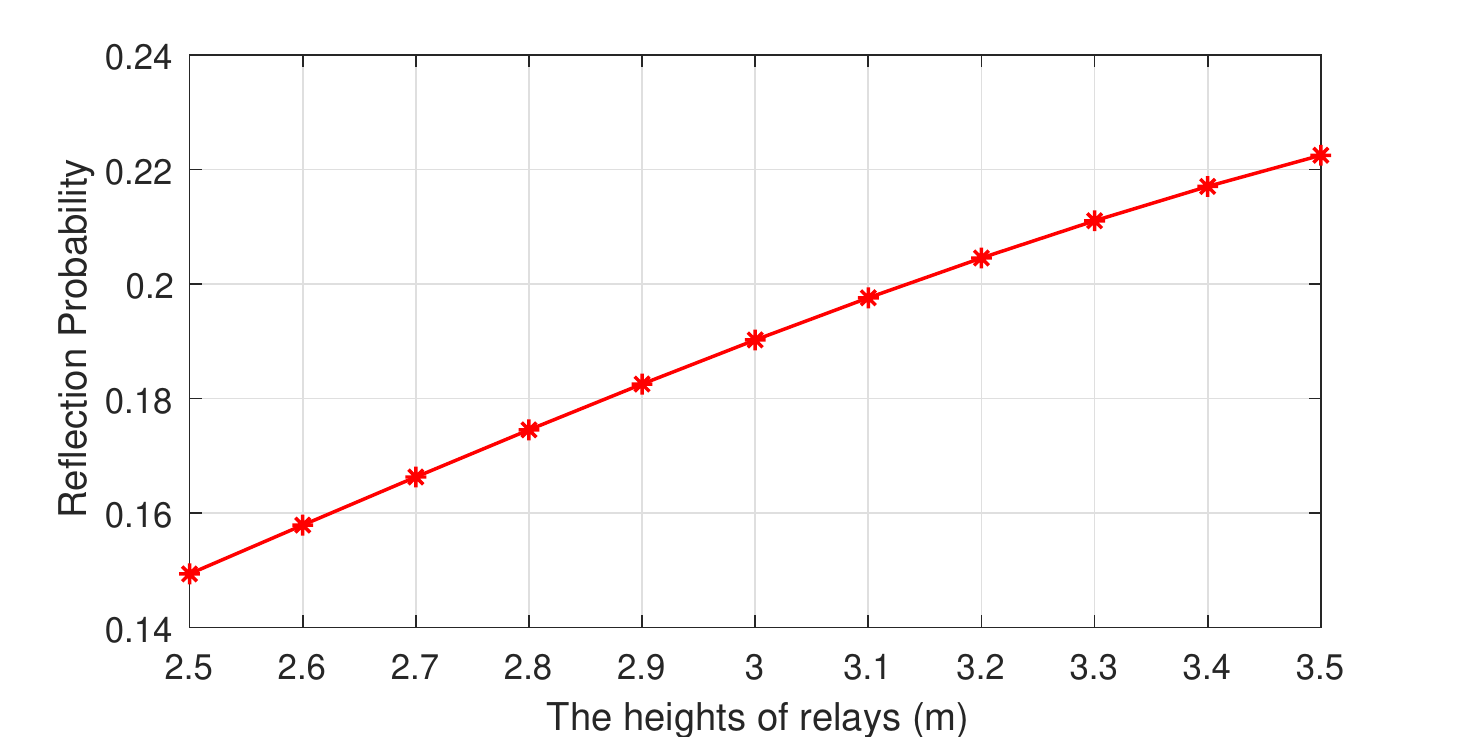}}
	\caption{Reflection probability vs. the height of relay nodes.}
	\label{Refl_Prob2}
\end{figure}

In addition, we investigate how the height of relay nodes in our topology affects the reflection probability. In Fig.~\ref{Refl_Prob2}, it is observed that the reflection probability will decrease if we decrease the height of deployed relays, for example, when vehicle's density is 8 * 10$^{-4}$, the reflection probability can be reduced to 15\% if we decrease the heights of relays from 3.5m to 2.5m.% however, the lower height of relays will result in a higher blockage probability, which should adopt HTPR algorithm for path reconfiguration.

As a result, we conclude that the secondary effects have only a small impact on SINR in our considered scenario, and the occurrence probability of reflection is examined not high with an usual vehicle's density on the road of an urban environment. 

%\section{Conclusion}
%
%In this work, we use both analytical method and simulations to justify that the interference caused secondary effects has only a very small impact on backhaul network performance in the IFTW topology. 

\vspace{+0.7cm}
\noindent \textbf{\large References}
\vspace{+0.2cm}

\noindent [1] J. Du, E. Onaran, et al., ``Gbps user rates using mmWave relayed backhaul with high gain antennas", \textit{IEEE Journal on Selected Areas in Communications}, 2017: 1363-1372.

\vspace{+0.1cm}
\noindent [2] Y. Liu and D. Blough, ``Analysis of Blockage Effects on Roadside Relay-assisted mmWave Backhaul Networks", \textit{Proc. of IEEE International Conference on Communications}, 2019.

\vspace{+0.1cm}
\noindent [3] Y. Liu, Q. Hu, and D. Blough, ``Blockage Avoidance in Relay Paths for Roadside mmWave Backhaul Networks", \textit{Proc. of IEEE Symposium on Personal, Indoor, and Mobile Radio Communications}, 2018.

\vspace{+0.1cm}
\noindent [4] M. Samimi, and T. Rappaport. ``Characterization of the 28 GHz millimeter-wave dense urban channel for future 5G mobile cellular", \textit{NYU Wireless TR} 1 (2014).

\vspace{+0.1cm}
\noindent [5] H. Zhao, et al. ``28 GHz millimeter wave cellular communication measurements for reflection and penetration loss in and around buildings in New York city", \textit{IEEE ICC}. 2013.

\vspace{+0.1cm}
\noindent [6] K. Sato, et al. ``Measurements of reflection and transmission characteristics of interior structures of office building in the 60-GHz band", \textit{IEEE Transactions on Antennas and Propagation} (1997): 1783-1792.

\vspace{+0.1cm}
\noindent [7] E. Violette, et al. ``Millimeter-wave propagation at street level in an urban environment", \textit{IEEE} \textit{Transactions on Geoscience and Remote Sensing} 26.3 (1988): 368-380.

\vspace{+0.1cm}
\noindent [8] Y. Liu, Q. Hu and D. Blough, ``Joint Link-level and Network-level Reconfiguration for mmWave Backhaul Survivability in Urban Environments", \textit{Proc. of Modeling, Analysis and Simulation of Wireless and Mobile Systems}, ACM, 2019.

\end{document}